\documentclass[twocolumn]{article}

\usepackage{arxiv}

\usepackage[all]{nowidow}

\usepackage[utf8]{inputenc} 
\usepackage[T1]{fontenc}    
\usepackage{hyperref}       
\usepackage{url}            
\usepackage{booktabs}       
\usepackage{amsfonts}       
\usepackage{nicefrac}       
\usepackage{microtype}      

\usepackage{float}  
\usepackage{graphicx}
\usepackage[labelfont=bf]{caption}
\usepackage{subcaption}     
\usepackage{tcolorbox}      

\title{Composer: Visual Cohort Analysis of Patient Outcomes}

\def \shortauthor {Rogers et al.}
\author{
  Jen Rogers \\
  University of Utah
     \And
   Nicholas Spina \\
   University of Utah 
      \And
   Ashley Neese \\
   University of Utah 
      \And
      Rachel Hess \\
   University of Utah 
         \And 
         Darrel Brodke \\
   University of Utah 
         \And
      Alexander Lex \\
  University of Utah \\
  \texttt{alex@sci.utah.edu}
}

\begin{document}
\twocolumn[
  \begin{@twocolumnfalse}
    \maketitle
    
\begin{abstract}
\textbf{Objective.}
Visual cohort analysis utilizing electronic health record (EHR) data has become an important tool in clinical assessment of patient outcomes. In this paper, we introduce Composer, a visual analysis tool for orthopedic surgeons to compare changes in physical functions of a patient cohort following various spinal procedures. The goal of our project is to help researchers analyze outcomes of procedures and facilitate informed decision-making about treatment options between patient and clinician. \\
\textbf{Methods.}
In collaboration with Orthopedic surgeons and researchers, we defined domain-specific user requirements to inform the design. We developed the tool in an iterative process with our collaborators to develop and refine functionality. With Composer, analysts can dynamically define a patient cohort using demographic information, clinical parameters, and events in patient medical histories and then analyze patient-reported outcome scores for the cohort over time, as well as compare it to other cohorts. Using Composer’s current iteration, we provide a usage scenario for use of the tool in a clinical setting. \\
\textbf{Conclusion.}
We have developed a prototype cohort analysis tool to help clinicians assess patient treatment options by analyzing prior cases with similar characteristics. Though Composer was designed using patient data specific to Orthopedic research, we believe the tool is generalizable to other healthcare domains. A long term goal for Composer is to develop the application into a shared decision making tool that allows translation of comparison and analysis from a clinician-facing interface into visual representations to communicate treatment options to patients. 
\end{abstract}
 
\vspace{2mm}    
\vspace{8mm}
  \end{@twocolumnfalse}
]

\begin{tcolorbox}[floatplacement=!b,float,left=2mm,right=2mm,top=1mm,bottom=1mm]
\small
This is the authors' preprint version of this paper. Please cite the following reference: \\
Jen Rogers, Nicholas Spina, Ashley Neese, Rachel Hess, Darrel Brodke, Alexander Lex. 
Composer: Visual Cohort Analysis of Patient Outcomes. 
\textit{Applied Clinical Informatics}, to appear, 2019.
\end{tcolorbox}

\section{Introduction}

Determining the best treatment option for patients with back pain involves an assessment of their medical histories and a comparison to similar patients. Such comparisons have relied on a physician’s memory of related prior cases, which can be influenced by cognitive biases. With an increasing amount of data available for patient populations in electronic health records (EHR), visual cohort analysis has gained attention as an informative analytic tool in healthcare. Recent work has shown the efficacy of using subsets of similar patients, referred to as cohorts, for outcome analysis and prediction in a “patient-like-me” approach~\cite{lee2015personalized, gallego2015bringing}. This approach can help clinicians assess treatment options for patients with certain characteristics or pre-existing conditions (comorbidities) that can influence recovery and response to treatment.
In this paper, we introduce Composer, a visual analysis tool for comparison of patient outcomes in cohorts under alternative treatment options. Composer was developed in collaboration with domain experts at the University of Utah Orthopedic Research Center. We incorporate outcome scores that are frequently measured over the course of treatment in the decision-making process, supplementing physicians’ memory of prior cases. We used the Patient Reported Outcomes Measurement Information System (PROMIS)~\cite{cella2010initial} scores as the metric for patient physical function and well-being over time.
The technical contributions of Composer include methods to flexibly define multiple patient cohorts based on EHR data and demographic attributes as well as medical codes associated with a given medical visit. We provide functionality for PROMIS score normalization to allow for alignment of score trajectories based on events in patient medical histories, such as surgery or injection. We also provide the ability to normalize scores from absolute measurements to relative change to identify improvement of patient physical function. Finally, we introduce aggregation methods to deal with larger patient cohorts.

\section{Background}
\paragraph{Cohort Analysis.} Most clinical guidelines are based on evidence from clinical trials and controlled studies. However, data collected from clinical trials, often sourced from a general population, may not provide an accurate reflection of potential outcomes for subsets of patients with pre-existing conditions and comorbidities~\cite{gallego2015bringing}. Clinicians are, therefore, interested in using EHR data and observational studies to better identify factors that can influence the recovery of such patients~\cite{thadhani2006cohort}. A cohort is defined as a subset of the general population that shares one or more defining characteristics. The analysis of cohorts has proven effective in the medical community for identifying factors that affect patient recovery and treatment.
In clinical applications, cohorts can be defined by utilizing patient data collected through the EHR system. The medical community has relied on cohort subsets sourced from a large body of EHR data that can be used for retrospective analysis~\cite{perer2015mining, thadhani2006cohort}. Cohorts of patients formed from EHR data have the potential to be used for “patients-like-me” comparisons~\cite{gallego2015bringing}, in which clinicians can define a cohort with attributes mirroring a given target patient. These comparisons can help identify factors that influence patient recovery and has been used to develop predictive tools that help domain experts determine the best treatment options for a given patient~\cite{du2017finding, gotz2014methodology, franklin2014treatmentexplorer}.

\paragraph{Patient-Reported Outcomes Measurement Information System.}
The Patient-Reported Outcomes Measurement Information System (PROMIS) is a validated measurement system that evaluates a range of patient physical functions~\cite{gruber2017validation}. In this paper, we use only PROMIS physical function (PF) scores. The PROMIS system defines the abilities of a patient with a specific score, which is determined by patient response to a series of questions~\cite{cella2010initial}. A patient who can run 10 miles without difficulty would have a PROMIS PF score of approximately 72, whereas a patient with a score of 32 can stand for a short period of time without difficulty~\cite{houck2017does}. If a patient has answered that they have trouble walking a mile, later questions will focus on a smaller range of physical abilities. The score system is converted to a t-score metric that ranges from 0 to 100, with an average ability score of 50 and a standard deviation of 10. All scores are scaled to values relative to the average score. For example, a score of 40 implies physical function that is one standard deviation lower than the score of the reference mean~\cite{cella2010initial}.
The University of Utah Orthopedic Research center has been a proponent in the use of PROMIS scores to assess patient outcomes~\cite{hung2014psychometric}. Recent research into PROMIS physical function scores to evaluate a given procedure relative to cost has identified PROMIS PF as a more accurate assessment of physical well-being for patients with spinal ailments than the Oswestry Disability Index, which is derived from patient reported questionnaire and is used to measure lower back pain. Due to its accuracy, PROMIS PF can be a valuable metric to evaluate patient well-being following treatment and assist in evidence-based decision-making for treatment options for patients with spinal conditions~\cite{brodke2017promis}.

\section{Domain Goals and Tasks}

This project emerged from a collaboration between two computer scientists with four medical researchers from the Orthopedic Research Center and the Department of Population Health Sciences at the University of Utah. The domain scientists are currently investigating the use of PROMIS scores as a measure of patient well-being and progression of physical function following various procedures for spinal ailments. In this project, we specifically target treatment options for Intervertebral Disc Herniation (IVDH). In meetings on a bi-weekly basis over 18 months, we collected notes on current EHR and PROMIS score use within the Orthopedic Research Center to identify domain goals and inform the design of our tool.
Two of the collaborators are spinal surgeons who have not used visualization of EHR data when considering a patient’s options for treatment. Instead, their assessments have been based on past experiences. When determining patient treatment options, they take into account demographics, medical comorbidities such as diabetes, prior treatments, and current symptoms and severity. They then choose the treatment that is likely to result in the best outcome while also considering other factors such as recovery time and cost. The main treatment options considered by our collaborators for patients suffering from IVDH are hemilaminectomy (a surgical procedure), steroid injection, and physical therapy, as well as combinations thereof. Because the medical histories and collected EHR data for the patient population are extensive and involve a variety of records and data types, we sought to develop a visual analysis solution that combines our collaborators’ data into a comprehensive dynamic interface that helps them identify trends in patient outcomes. We identified three functionality requirements that inform the design of Composer, defined below:

\textbf{R1.} Define meaningful cohorts of patients and analyze how this subset of patients reacts to various treatments and procedures. The clinicians need to be able to form cohorts from the EHR data based on patient demographic information, treatment history, medical records, and initial physical function scores.

\textbf{R2.} Compare the outcomes of different cohorts, for example, physical function outcomes following different treatment options in otherwise identical cohorts, or to identify an effect of a comorbidity.

\textbf{R3.} Normalize Physical Function Scores in several ways to successfully analyze and compare cohort outcomes, following an event, such as surgery.
\vspace{-.5mm}    
\section{Related Work}
Visualization of patterns in patient medical histories helps identify risk factors that influence patient recovery following treatment~\cite{wang2008aligning}. Recently developed clinical tools provide visual support for users, often in the form of aggregated representations of patient data derived from EHR as well as visual comparisons for patient outcomes and trajectories~\cite{bernard2015visual, gotz2014methodology, perer2015mining}. Composer is related to various tools and techniques for cohort definition and EHR analysis, which we discuss below.
\vspace{-.5mm}     
\paragraph{Cohort Definition.} Cohort definition is a vital first step for analysis. Emergent patterns identified in cohort behavior and outcome remain dependent on the accuracy of the cohort creation~\cite{krause2016supporting} and therefore, cohort definition tools often provide visual feedback to track stages in cohort definition~\cite{krause2016supporting}. We included a visual representation of each filter layer for a cohort in Composer and have extended this idea to allow dynamic changes to filters.
\paragraph{Cohort Comparison.} Current visual tools often provide users the ability to compare clinical pathways and outcomes of patients. These comparisons help users identify differences in patient outcomes between two defined cohorts and diverging event sequences within a given cohort’s records~\cite{perer2015mining}. Normalization to a standard time metric and alignment at events in the patient histories facilitate comparison and highlight patterns within the data~\cite{wang2008aligning}. This time metric, often in the form of days or visits, allows patient histories to be viewed along a common axis. A tool by Bernard et al.~\cite{bernard2015visual} allows realignment of events, e.g., when metastases develop in cancer patients. By sorting and realigning, users can better see trends between events and their corresponding phases. Comparisons can be used for identifying both significant differences as well as similarities and recurring patterns. In contrast to Bernard et al., Composer represents patient trajectories as single lines layered over one another which allows visualization of a larger number of patient trajectories at once. In Composer, we normalize patient data to a standard day metric and allow users to realign scores to a common procedure event. This facilitates comparison of score fluctuation for cohorts containing several hundred patients after given events by viewing patient score change aligned on a common axis.

\paragraph{Aggregation.} Much patient data includes event sequences and temporal information. With a large amount of patient data over a span of years, visualization of patient care pathways and events can prove difficult. Clinicians must be able to identify patterns of events within a single patient’s medical history and recurring trends between multiple patients’ records~\cite{monroe2013temporal}. Data, therefore, are often aggregated and summarized to identify emergent patterns within the cohort’s medical timelines and track progression~\cite{widanagamaachchi2017interactive}. Aggregation can help with pattern identification within complex temporal data by reducing the visual complexity, although it can also hide subtle trends in the data~\cite{monroe2013temporal, gleicher2018considerations}.  Composer uses aggregation of individual scores to show emergent trends in PROMIS score fluctuation without the clutter of hundreds of individual plotted trajectories of patient scores at once. Users can view the scores individually or aggregated at their discretion. 

\paragraph{Making Relationships in the Data Explicit.} Many recent tools facilitate cohort definition and analysis by making relationships between events and static attributes more explicit. Bernard et al.’s visual analysis tool for patients with prostate cancer visualizes distributions of static attributes in the data and indicates when an attribute’s frequency is higher or lower in the cohort relative to the population. This visual information is valuable to the domain expert as it provides insight into filter constraints on attributes that might have influenced a subset of patient outcomes~\cite{bernard2015visual}. Du et al.’s EventAction is a prescriptive visual tool for event sequences. It provides plots showing positive and negative correlations between categories and outcomes~\cite{du2016eventaction}. Another method of highlighting significant relationships within the cohort data is through visual hierarchy and color. Many visual tools provide color coded highlighting to emphasize significant events~\cite{gschwandtner2011design, bernard2015visual}. By making these relationships explicit, users can make informed decisions to determine the next steps. We have incorporated these methods in Composer by providing distribution plots to show the number of patients in the entire population who meet the requirements for each filter category. For example, users can see the distribution spread of patient BMI measurements. We also provide visual representation of each filter constraint on a given cohort along with the number of patients at each filter stage.

\section{Composer Design}
\label{sec:design}

\begin{figure*}[t]
\centering
\includegraphics[width=\linewidth]{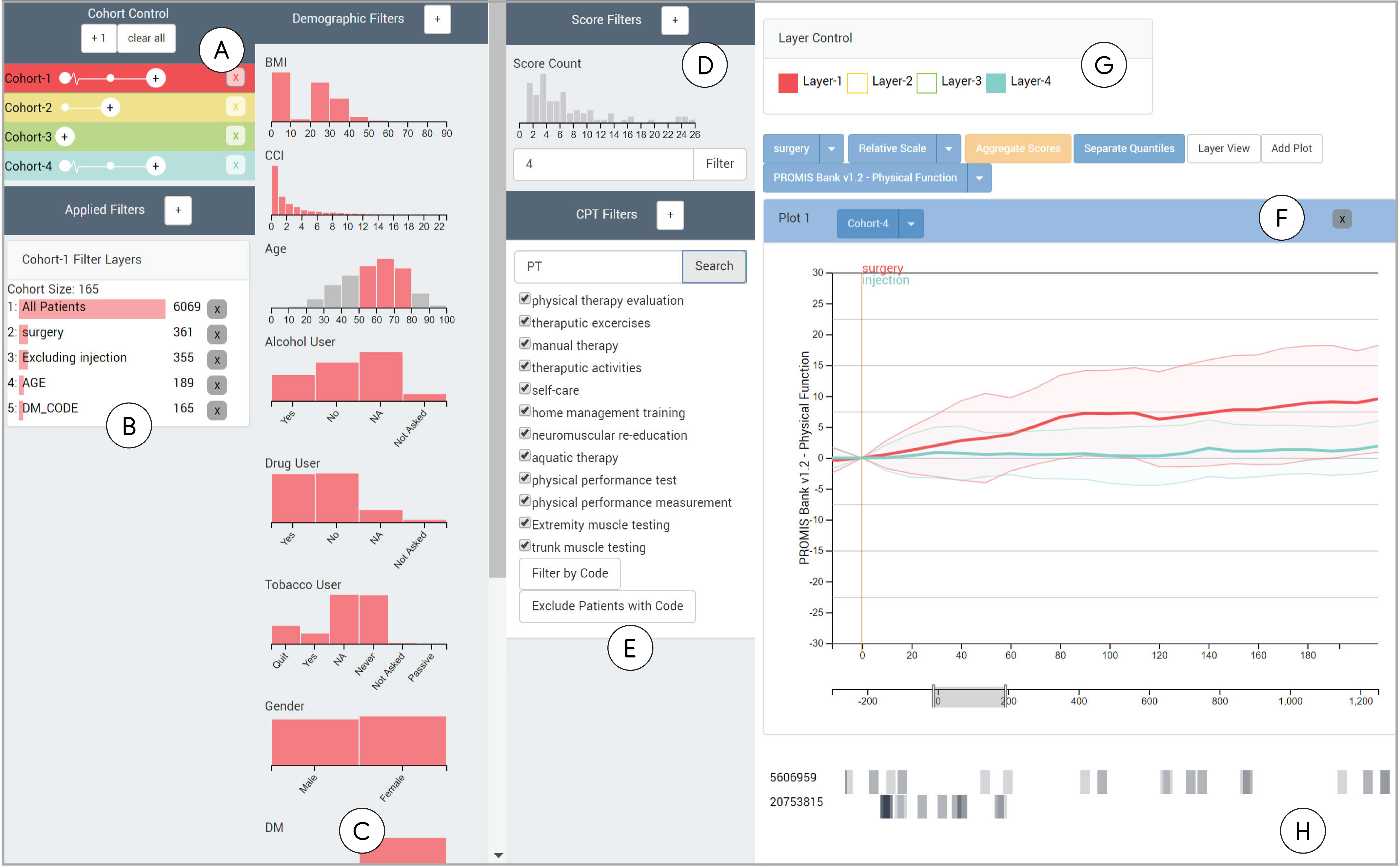}
\caption{Composer Overview. Composer consists of interfaces for flexibly defining cohorts, and for displaying the physical function scores of patients treated for back problems over time in these cohorts. (A) Patient cohorts can be added and branched in the cohort control interface . (B) A history of all filters applied to the selected cohort is shown below.  Cohorts can be defined using (C) filters applied to demographic information, (D) recorded score frequencies , (E) and presence or absence of procedural codes. (F) The main interface is a chart showing either individual lines, or aggregated areas. A zero-point for the PROMIS scores, indicated by the horizontal red line, can be flexibly defined to align all patients by a specific event, such as a medical intervention. (G) The layer panel provides the ability to hide layers corresponding to the cohorts. (E) Users can select individual patient lines to show orders associated with their medical records in the timeframe specified in the timeline below the main plot. Selected patients are identified by their patient id, shown on the left hand side of the event line.}
\vspace{-3mm}
\label{fig:composer_overview}
\end{figure*}

Composer, shown in Figure~\ref{fig:composer_overview}, consists of two components: the cohort definition interface, and the visualization of PROMIS physical function scores. The cohort definition interface is contained within the collapsible sidebars on the left, while the outcome score interface is placed on the right. We chose to encode the score trajectories as a line plot, similar to the style of chart our collaborators currently use to represent PROMIS score trajectories, as this is both, perceptually efficient and a common representation to view change in a metric over a period of time.

\subsection{Cohort Creation}

Our collaborators need the ability to define a cohort from a set of specific attributes and medical histories (R1). In Composer’s filter sidebar (see Figure~\ref{fig:composer_overview} A-E) cohorts can be defined by demographic information such as age or gender, in addition to other factors deemed relevant, like smoking habits. The filter sidebar is divided into Demographic, Score, and CPT (Current Procedural Terminology; codes used to identify procedures) sections. Within the demographic filters, we use histograms to visualize the distributions of attributes in the patient population (Figure~\ref{fig:composer_overview}C). The histograms also serve as means to interact with a filter through brushing for quantitative attributes and selections for categorical ones. In addition to demographic variables, cohorts can also be defined by the number of recorded PROMIS scores for a patient (Figure~\ref{fig:composer_overview}D), or based on the presence or absence of procedure codes in patient histories (Figure~\ref{fig:composer_overview}E). This allows analysts to, for example, separate patients that have received a specific surgery from those who have not. With each cohort refinement, a filter layer is added to the sidebar as a visual history of filters used and cohort size at the given filter (Figure~\ref{fig:composer_overview}B). Individual filters and cohorts can be removed from the filter history or updated at any time in the cohort sidebar (Figure~\ref{fig:composer_overview}A). Composer enables analysts to define multiple cohorts simultaneously. Each cohort is represented as a colored bar and assigned a unique label and color, which is kept consistent across the interface. Within the bar, filters are represented as white nodes. If more than three filters are present, they are aggregated. 
To facilitate cohort comparison (R2), cohorts can be branched. Once branched, the filter constraints of the parent cohort are duplicated in the branch but can be refined independently. This allows users  to add diverging filters for an attribute that an analyst believes may influence the outcome of a treatment. For example, users may want to see if there is a difference in patient trajectories after physical therapy, if they have also had a steroid injection. To do that, they can define an initial cohort, branch it, and apply filters for subsequent steroid injections vs no injections to the branches. 

\begin{figure*}[!ht]
  \begin{subfigure}[b]{.33\linewidth}
    \centering
  \includegraphics[width=\textwidth]{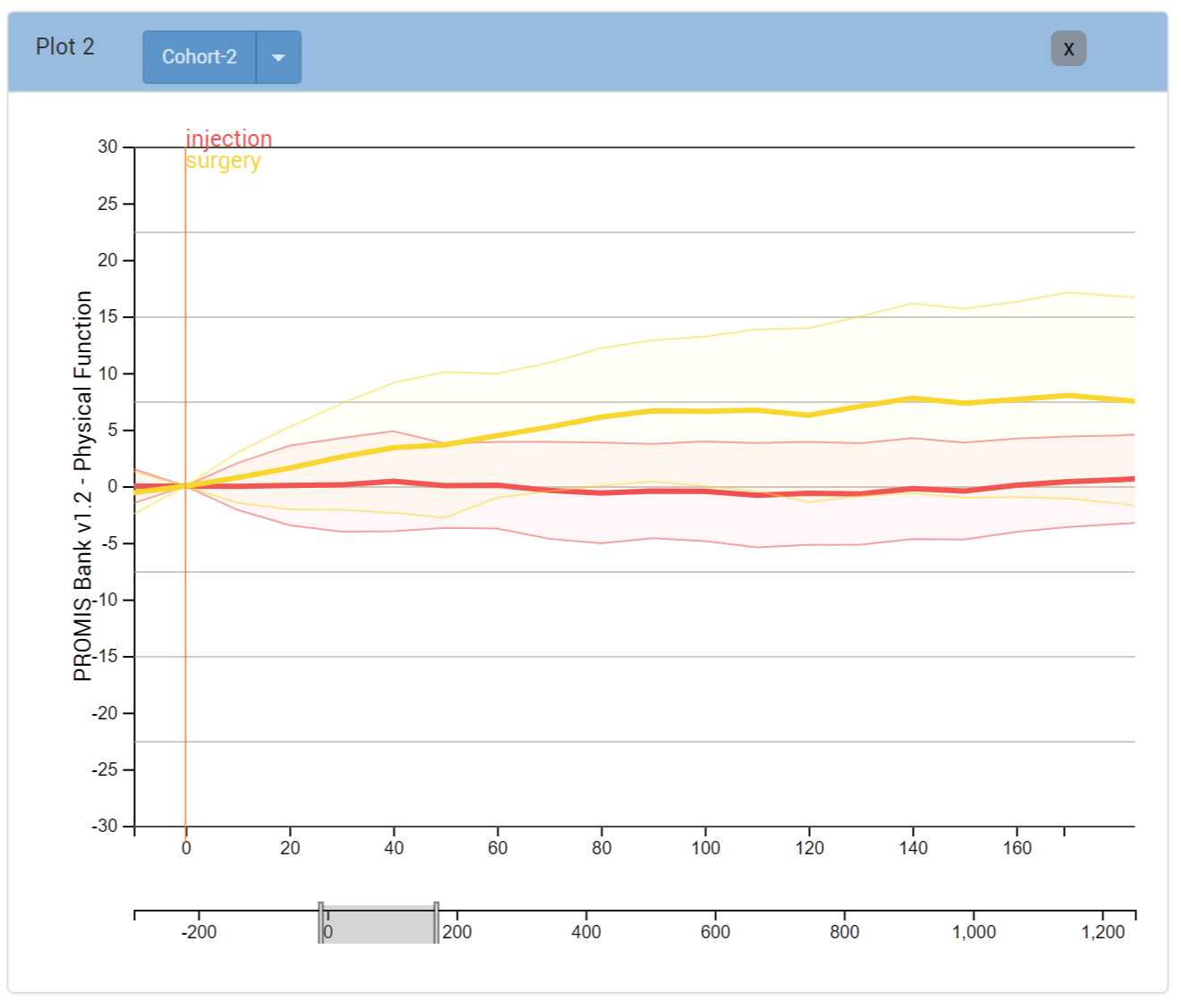}
   \caption{Layering.}
    \label{fig:comparison_layering}
  \end{subfigure}%
  \hfill
  \begin{subfigure}[b]{.66\linewidth}
    \centering
    \includegraphics[width=\textwidth]{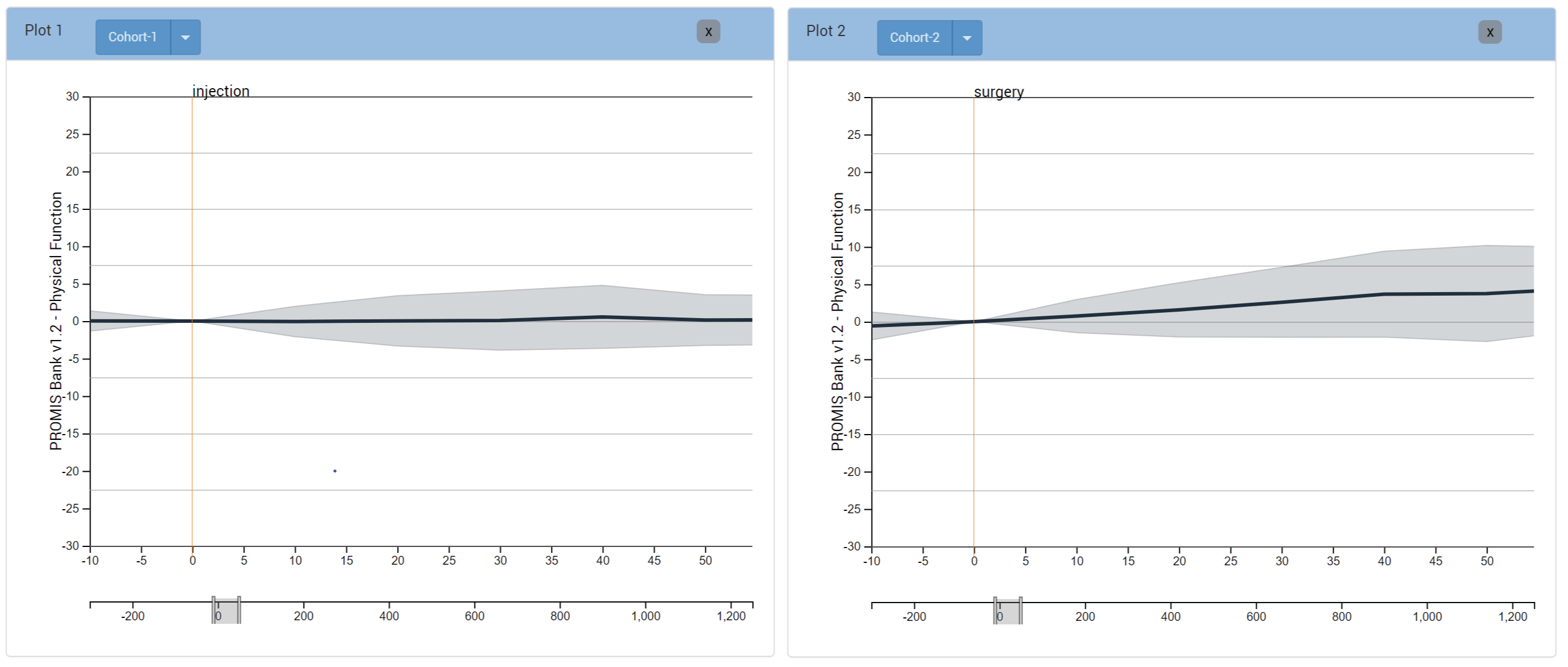}
    \subcaption{Juxtaposition.}
    \label{fig:comparision_juxtaposition}
  \end{subfigure}
  \caption{Differences in PROMIS scores after surgery and injection compared by (a) layering and (b) juxtaposition of multiple plots. Both methods allow for comparison of score change after different treatment events.}
  \label{fig:comparison}
\end{figure*}

\subsection{ Outcome Score Comparison}
PROMIS physical function scores for the defined cohort are visualized as individual lines showing the course of physical function for each patient over time. The time-window can be resized as desired. By default, we align by the first PROMIS score, yet alignment by a specific clinical event, such as surgery or the start of physical therapy, are often more informative. When different cohorts are aligned by different events this way, the relative progression after the event can be evaluated. This facilitates comparison between cohorts (R2) by allowing the user to manipulate the alignment and scale in a dynamic way (R3). We use juxtaposition and superimposition to compare between cohorts18, which have different trade-offs as far as required display space and clutter in a single plot are concerned. Juxtaposition  allows users to add multiple plots to evaluate cohort trajectories in a side-by-side comparison (Figure~\ref{fig:comparison}). Superimposition shows different layers on top of each other (Figure~\ref{fig:composer_overview}F). We allow analysts to toggle layers individually (Figure~\ref{fig:composer_overview}G).

\begin{figure*}[t]
  \begin{subfigure}[b]{.45\linewidth}
    \centering
  \includegraphics[width=\textwidth]{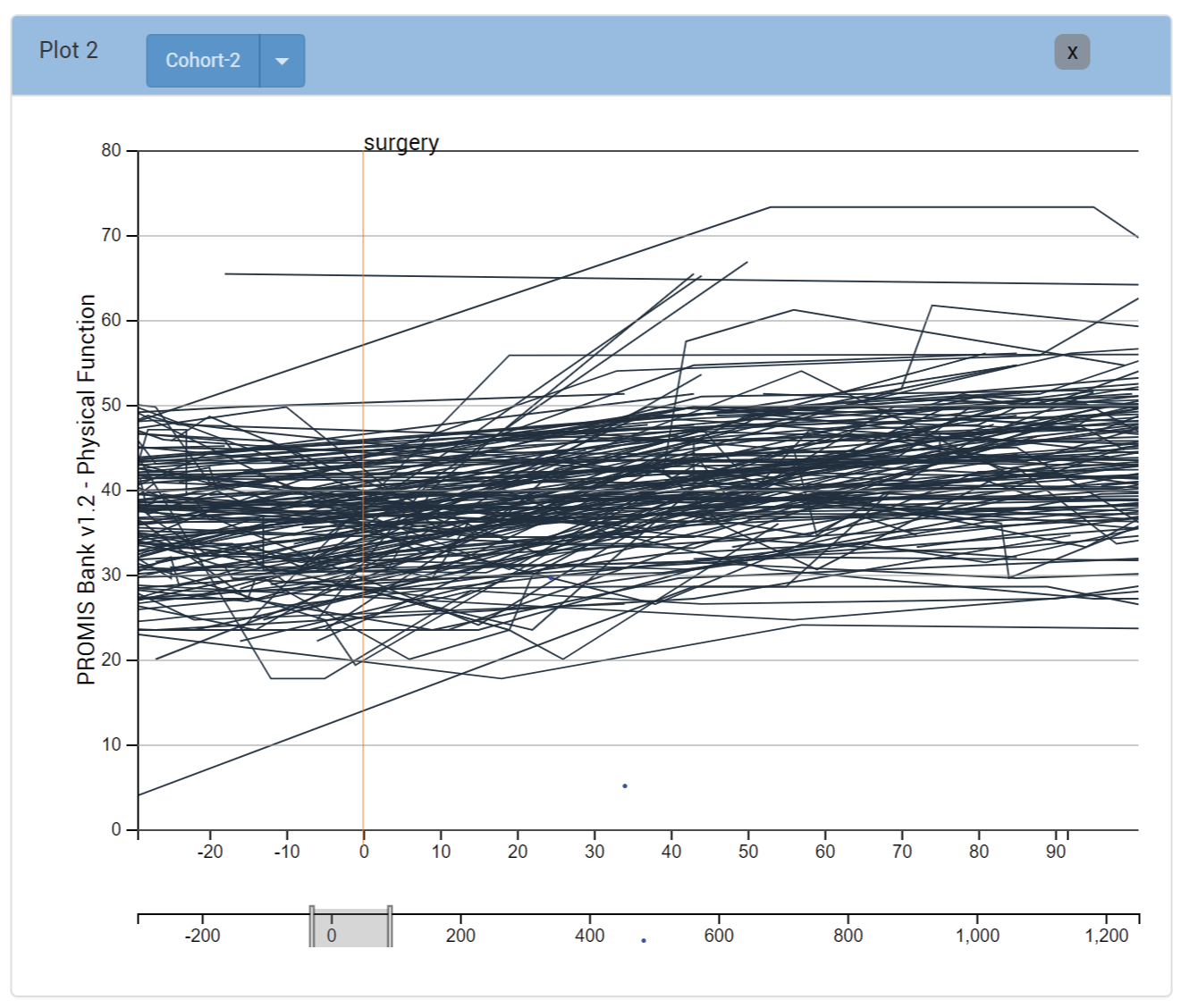}
   \caption{Absolute Scaling.}
\label{fig:scale-absolute}
  \end{subfigure}%
  \hfill
  \begin{subfigure}[b]{.45\linewidth}
    \centering
    \includegraphics[width=\textwidth]{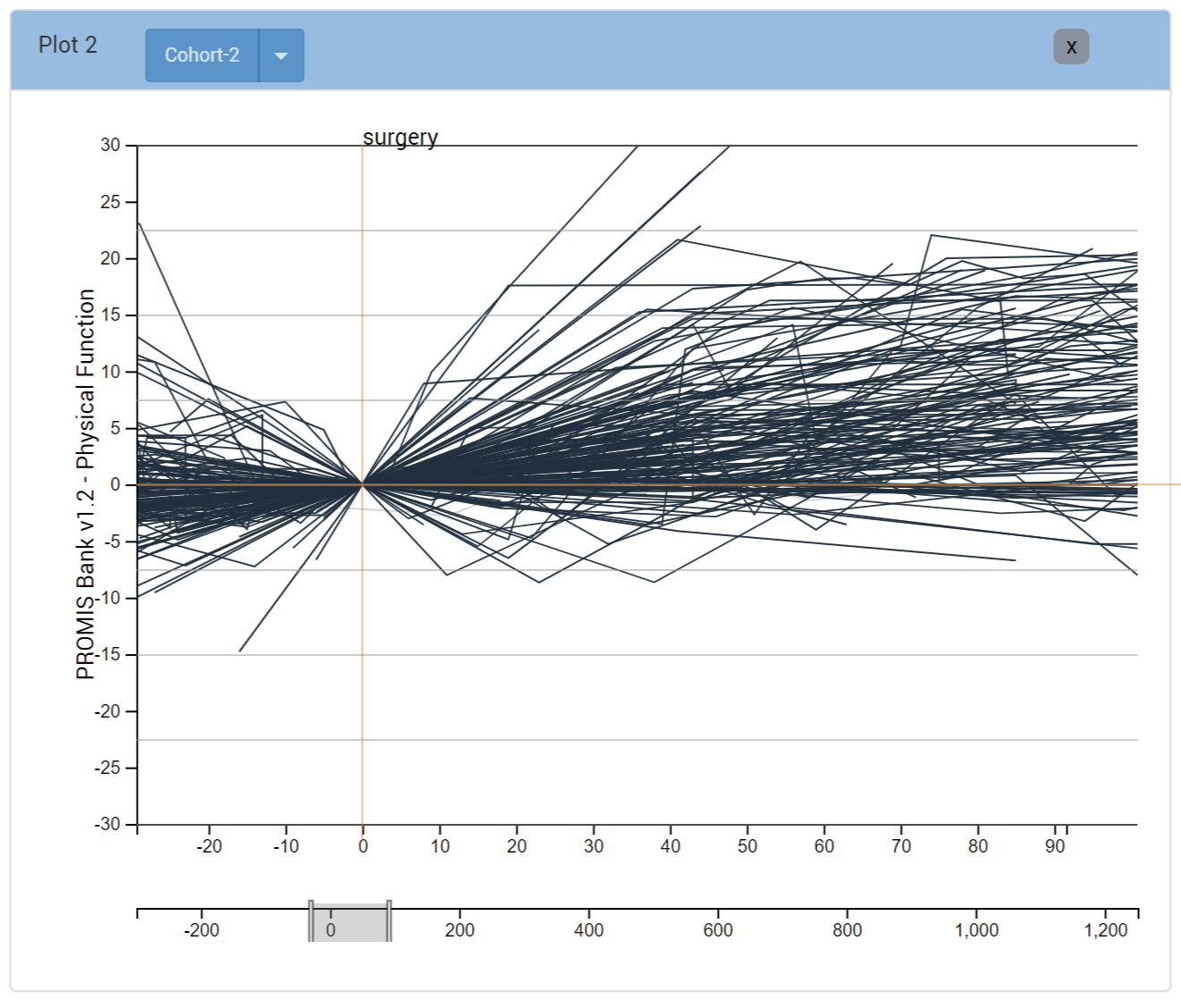}
    \subcaption{Relative Scaling.}
 \label{fig:scale-relative}
  \end{subfigure}
  \caption{View of score plots using (a) absolute and (b) relative scales. Each line represents an individual patient. Relative scales show change in PROMIS PF score, calculated from the score at the day zero event. In this case the patient score trajectories are aligned by the day of surgery. With a larger cohort, the general trend for patient progression can be difficult to see, which we address by providing aggregation functionality.}
  \label{fig:scale}
\end{figure*}

\vspace{-.5mm}    
\paragraph{Dynamic Score Scales and Normalization.} The physical function scores used by the domain experts are often subtle in absolute measured change (see Figure~\ref{fig:scale-absolute}), yet these subtle changes often have significant impact on the perceived well-being of patients. Change in patient scores are further obscured as patients in the same cohort have different baseline scores. To emphasize change and normalize the baseline, analysts can view scores on a normalized scale that visualizes relative score change for the patients, as shown in Figure~\ref{fig:scale-relative}. With the option of both absolute and relative score scales, analysts can assess the cohort’s overall trend in baseline score measurements as well as trends in score fluctuation. By showing relative score change and making the relationship between cohort scores more explicit, analysts can see differences in outcome trajectories during comparison more clearly. In addition, users have the ability to adjust the timeframe of the line chart. The timeframe is specified through brushing a selection of the lower timeline that extends the minimum and maximum range of days for all patient records  (See Figure~\ref{fig:scale}).

\begin{figure*}[t]
  \begin{subfigure}[b]{.45\linewidth}
    \centering
  \includegraphics[width=\textwidth]{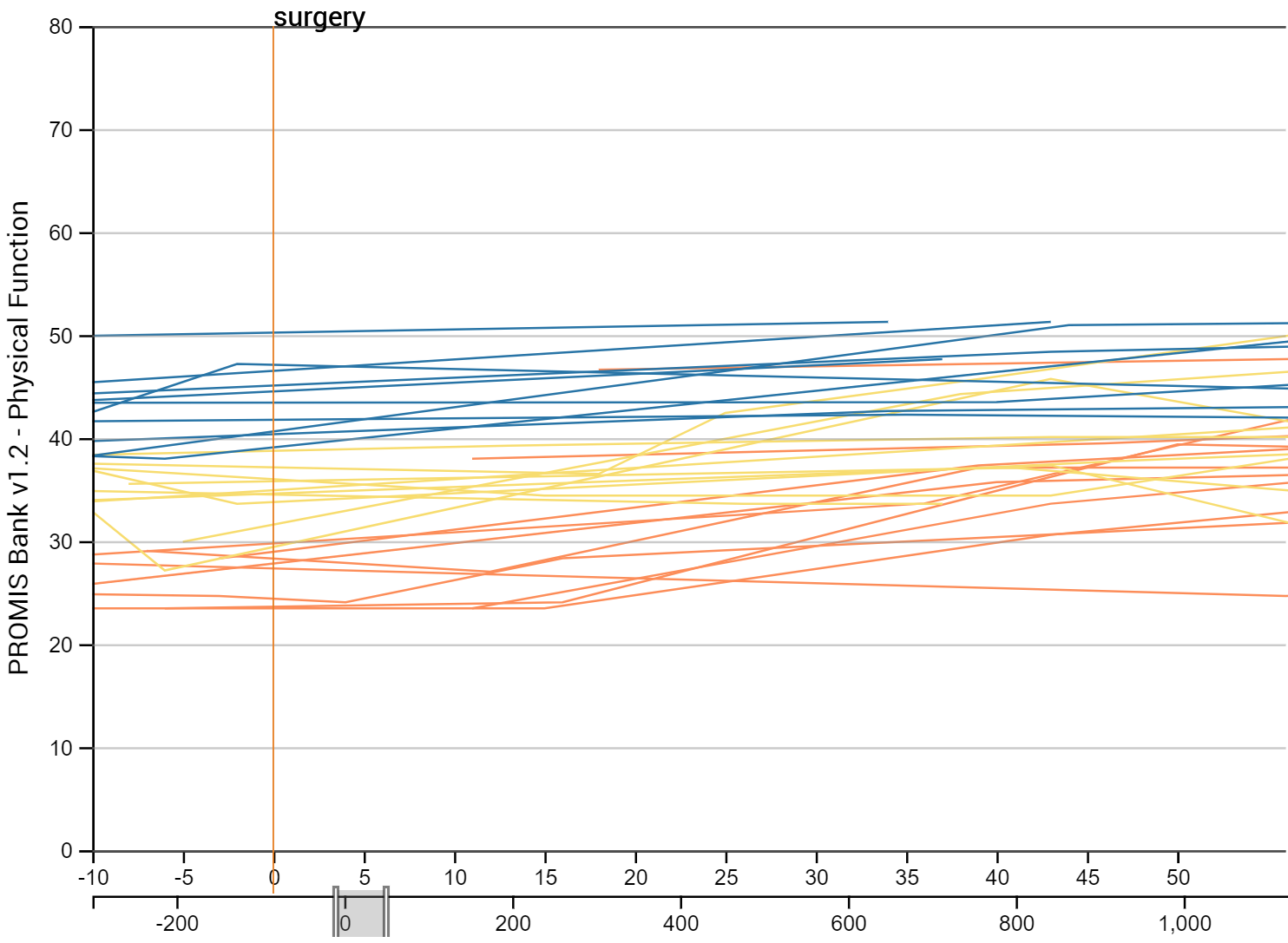}
   \caption{Color coded quantiles.}
\label{fig:quantiles-color-coded}
  \end{subfigure}%
  \hfill
  \begin{subfigure}[b]{.45\linewidth}
    \centering
    \includegraphics[width=\textwidth]{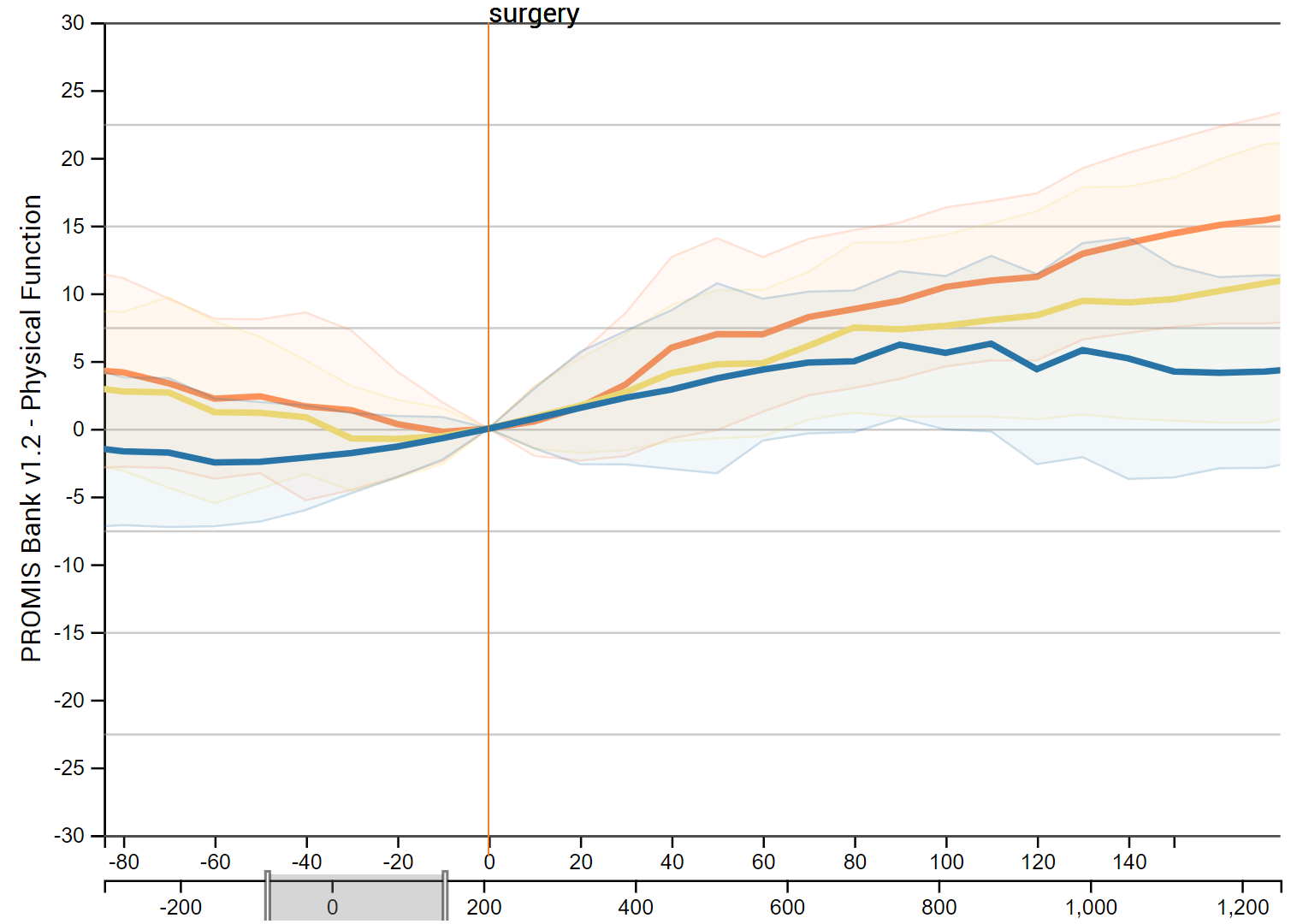}
    \subcaption{Aggregated quantiles.}
 \label{fig:quantiles-aggregated}
  \end{subfigure}
  \caption{View of patient scores separated and color coded by quantiles. The PROMIS PF scores were separated into quartiles, shown as individual lines in (a) and aggregated area charts in (b). The orange marks represents the top quartile, the yellow marks the inter-quartile range, and the blue marks the bottom quartile.}
  \label{fig:quantiles}
\end{figure*}

\paragraph{Separation of Scores by Quantiles.} Even in a well-defined cohort, patient outcomes can be markedly different. Due to this heterogeneity, our collaborators need the ability to separate the cohort into quantiles that communicate how, for example, the physical function changes for the top 25 percent of patients in the cohort (Figure~\ref{fig:quantiles-color-coded}). In Composer, a cohort can be divided by quartiles. We calculate these quartiles by the average change in score over a user-adjustable period of days following a given event.

\paragraph{Aggregation of Scores.} Frequently, our collaborators do not need to view individual patients, but rather are interested in aggregate representation of scores. To address this need, we provide means to aggregate the scores of a cohort to visualize the interquartile range with a line representing the median. Aggregated cohort scores can also be separated by quantiles to more clearly identify any difference in score change within subsets of the cohort that have different baseline measurements, as shown in Figure~\ref{fig:quantiles-aggregated}.	
 \paragraph{Individual Patient CPT History View.} For further analysis of procedure code distributions and procedure frequency, analysts can select an individual patient from a group of patient trajectories in the score chart to view all orders associated with that patient’s medical history (see Figure~\ref{fig:composer_overview}H). These histories are cropped to the timeframe specified in the score chart and aligned with its timeline. For example, if the score chart shows trajectories between 20 days before an injection and 60 days after, the individual timeline would reflect the same timeframe. These events can provide context for individual cases, but can also be used to further filter a cohort. Analysts can view patient histories by selecting the patient’s PROMIS scores on a given plot. The events then appear below the plot, aligned on the same time.
 
 \subsection{Implementation}
 Composer is open source and was developed with Typescript using the D3.js library for visualization. The prototype is a Phovea client/server application21. The Code for Composer can be found at https://github.com/visdesignlab/Composer. Data used for development and to inform the usage scenario was sourced from a sample of EHR provided by our collaborators from the Orthopedic Research Center’s database and was preprocessed in Python.
 
 \section{Usage Scenario}
 Here we describe a usage scenario to illustrate a typical use case for composer as it can be used by our domain collaborators. 
A surgeon sees a patient suffering from a herniated disc. While evaluating potential treatment options for the patient, she defines a cohort in Composer using constraints based on the given patient’s medical history. She filters by the patient’s age range, specifies the cohort to only include diabetic patients, and filters just those patients that have had physical therapy evaluation. The cohort defined by these patient specific filters contains 3317 patients. She branches the cohort and filters the initial branch by those that have had surgery, but have not had an injection. She then filters the secondary branch by those patients that have had an injection but not surgery. Aligning each cohort by the surgery or injection event they were filtered by, she can view the diverging cohorts superimposed over one another and visually compare differences in PROMIS PF score fluctuation between the two. She can then aggregate the individual scores to show only the median PROMIS score within the cohort. Next, she normalizes the scores from the absolute score measurement to relative score change, so that she can visually compare the difference in score change between the two to determine what treatment appears to produce better outcomes (Figure~\ref{fig:comparison}). After comparing the change in score across a span of 150 days after treatment, she can see that surgery had a greater positive change in physical function, which is clearly visible after the first month (Figure~\ref{fig:comparison}). She can take this into consideration when determining patient treatment options, and show this visualization to the patient when discussing treatment options.

\section{Discussion and Limitations }

Composer is under active development, with progressive iterations being made in response to feedback received from meetings with collaborators.
\paragraph{Evaluation.} We considered various strategies to evaluate our contribution, including collecting feedback from our collaborators, and comparing to other tools. While we have received positive feedback from our collaborators, we chose to not report it in detail due to the potential for biases. Ultimately, we have chose to validate Composer through a usage scenario and the careful justification of our design decisions, which are accepted practices in user-centered design~\cite{greenberg2008usability}. However, the larger question is whether using a tool like composer will lead to better outcomes. We are currently planning a longitudinal study using the tool and measure provider and patient satisfaction, but also outcomes. However, such a study is beyond the scope of this paper. 
\paragraph{Data Integration.} Currently, the data used in Composer is a large but static dataset of patients pulled from the Orthopedic Center’s database. By using a static snapshot, we have full control over processing and data manipulation for initial development while avoiding issues such as permissions and compatibility associated with a deep integration with the EHR system. We expect to be able to run a longitudinal evaluation without integrating Composers, however, this creates manual effort when incorporating new patient data or updating existing data.  As we develop Composer beyond its proof-of-concept stage and past a formal evaluation, we intend to integrate the tool with our collaborator’s EHR system. 
\paragraph{Data Cleanup.} A challenge common to systems operating on data extracted from electronic health records is the data’s messiness and inconsistency. We address sparse outcome scores by interpolation, yet we acknowledge the limitation in accuracy for interpolated patient trajectories for those patients that have lower score frequencies. We exclude patients with fewer than three PROMIS PF scores. We also do not currently consider systematic biases in score trajectory: for example, it is likely that we have less data for patients with good outcomes, as they do not come for follow-ups. We hope to mitigate these limitations in future iterations of the tool by making uncertainty in patient trajectories more explicit in visualization and statistical representation. 
\vspace{-.5mm}    
\section{Conclusion Future Work}
\vspace{-.5mm}    
In this paper, we outlined the domain analysis for and the design of Composer, an application to visualize and compare patient cohorts and their physical function trajectories. This tool was developed in collaboration with domain experts from the Orthopedic Research Center at the University of Utah, with their current research in the efficacy of PROMIS scores to evaluate physical function of patients with lower back conditions. Immediate development of the tool will focus on addressing the limitations described in the previous section. In the near future, we plan to provide a more extensive statistical breakdown of cohort medical history with the inclusion of ICD codes. As distributions of events and attributes become more explicit, users will be able to apply more accurate filtering constraints to define cohorts. Additionally, we plan to provide more control of the CPT filter codes as they appear within the patient record, and inclusion of sequence specific event filters. As recent literature has shown, medical event sequences can provide important clues on patient outcomes~\cite{franklin2014treatmentexplorer, perer2015mining, du2016eventaction}. Currently, target patient outcomes are interpreted implicitly by evaluating score trajectories of a body of similar patients. We intend to improve interpretation of target patient outcomes through explicit data-driven forecasting of score trajectories using a larger patient sample, informed by previous work from Buono et al.~\cite{buono2007similarity}. Composer’s initial development targets Orthopedic patient comparisons and evaluation, we expect to be able to generalize it to other cases where outcome measures over time are the subject of the analysis. We also anticipate that our cohort definition interface could be applied in an even broader context. 
The long term goal for Composer is the addition of an interface for shared decision making in which insight from exploration in the current interface could be translated into visualizations that would facilitate the explanation of treatment choices and potential outcomes to the patient, and the integration of other measures, such as cost. As previously mentioned, we also plan a clinical evaluation of the tool.

\vspace{-5mm}
\section*{Acknowledgements}
\vspace{-5mm}
We thank Sahar Mehrpour for initial data analysis and implementation. This project is funded by the Orthopaedic Research Center and NSF IIS 1751238.

\bibliographystyle{unsrt}  

\bibliography{references}  

\end{document}